\documentclass[aps,showpacs,superscriptaddress,preprint]{revtex4}
\usepackage{amsmath}
\usepackage{graphicx}
\usepackage{hyperref}
\usepackage{multirow}
\newcommand{\GeV}{\mbox{GeV}}
\makeatother

\begin{document}

\title{Fourth generation  Majorana neutrino, dark matter and Higgs physics}
\author{Shou-Shan Bao}
\email{ssbao@sdu.edu.cn}
\affiliation{School of Physics, Shandong University, Jinan, 250100, P. R. China}
\author{Xue Gong}
\email{gongxue@mail.sdu.edu.cn}
\affiliation{School of Physics, Shandong University, Jinan, 250100, P. R. China}
\author{Zong-Guo Si}
\email{zgsi@sdu.edu.cn}
\affiliation{School of Physics, Shandong University, Jinan, 250100, P. R. China}
\affiliation{Center for High-Energy Physics, Peking University, Beijing 100871, P. R. China}

\author{Yu-Feng Zhou}
\email{yfzhou@itp.ac.cn}
\affiliation{State Key Laboratory of Theoretical Physics,\\ Kavli Institute for Theoretical Physics China, Institute of Theoretical Physics, Chinese Academy of Sciences, Beijing, 100190, P. R. China}

\begin{abstract}
We consider extensions of the standard model with fourth generation fermions (SM4) in which
extra  symmetries are introduced such that the transitions between the fourth generation fermions and
the ones in the first three generations are forbidden. In these models, the stringent lower bounds on the masses of fourth generation quarks from direct searches are relaxed, and the lightest fourth neutrino is allowed to be stable
and light enough to trigger the Higgs boson invisible decay.
In addition, the fourth Majorana neutrino can be a subdominant but highly detectable dark matter component.
We perform a global analysis of the current LHC data on
the Higgs production and decay in this type of SM4.
The results show that the mass of the lightest fourth Majorana neutrino  is confined in the range $\sim 41-59$ GeV.
Within the allowed parameter space, the predicted effective cross-section for
spin-independent DM-nucleus scattering is
$\sim 3\times 10^{-48}-6\times 10^{-46} \text{ cm}^{2}$,
which is close to the current Xenon100 upper limit and is within the reach of the Xenon1T experiment in the near future.
The predicted spin-dependent cross sections can also reach $\sim 8\times 10^{-40}\text{ cm}^{2}$.

\end{abstract}

\maketitle
\section{Introduction}

Models with chiral fourth generation fermions (SM4) are
well-motivated extensions of the standard model (SM),
and have been studied extensively in the literature (for reviews, see e.g.~\cite{Frampton:1999xi,Holdom:2009rf}).
The condition for $CP$ symmetry violation in the SM only requires the existence of at least
three generations of chiral fermions~\cite{Kobayashi:1973fv}.
There is however no upper limit on the number of generations from the first principle.
In the SM, the amount of $CP$ violation is not large enough to
explain the baryon-antibaryon asymmetry in the Universe.
The SM also fails to provide a valid dark matter (DM) candidate.
In some non-minimal models of SM4, extra left- and right-hand neutrinos are
introduced, these  neutrinos  can be detectable dark matter particles.
With massive quarks in the fourth generation,
it has been proposed that the electroweak symmetry breaking can be a dynamical feature of the SM~\cite{Holdom:1986rn,Carpenter:1989ij,Hill:1990ge,Hung:2009hy}.

The simplest version of SM4 contains  a sequential fourth generation of fermions.
The fourth neutrino can be  either Dirac or Majorana. This simple model is already stringently constrained by  various experiments. The current lower bound on the mass of the unstable fourth generation charged lepton $e_4$ is $m_{e4}\geq 100.8$ GeV from the search for the decay $e_4\to \nu_{4} W^{-}$ where $\nu_4$ is the fourth neutrino~\cite{Beringer:1900zz}.
The LEP-II data have shown that the number of light active neutrinos is three. From the invisible width of $Z$ boson, the lower bound for the mass of an unstable Dirac neutrino  is set to be $m_{\nu4}\geq 101.3$ GeV from the decay $\nu_4 \to e^-W^{+}$. The lower bound for a stable Dirac (Majorana) neutrino is  roughly half of the $Z$ boson mass, i.e., $m_{\nu_{4}}> 45~(39.5)$ GeV~\cite{Beringer:1900zz}.

Direct searches for fourth generation quarks at the Tevatron  and the LHC  can
push the lower limits on  the masses of  the fourth generation quarks close to the non-perturbative region,
which however depends strongly on  assumptions.
For instance,
the searches for the fourth down-type quarks $b^\prime\bar{b}^\prime$ pair-production set a lower limit of
$m_{b^\prime}>611$ GeV  by CMS~\cite{Chatrchyan:2012yea}
and $m_{b^\prime}> 480$ GeV by ATLAS~\cite{ATLAS:2012aw},
assuming $\mbox{Br}(b^\prime\to t W)=1$;
through searching for the single production $p\bar{p}\to b^\prime q$,
the limit is found to be  $m_{b^\prime}> 430$ GeV
(assuming $\mbox{Br}(b^\prime\to d Z)=1$) or
$m_{b^\prime}> 693$ GeV (assuming $\mbox{Br}(b^\prime\to u W)=1$) by D0~\cite{Abazov:2010ku};
through searching for the fourth up-type quark $t^\prime\bar{t}^\prime$ pair-production,
assuming subsequent decay $t^\prime\to bW$,
the obtained bound is $m_{t^\prime}>570$ GeV by CMS~\cite{Chatrchyan:2012vu} and
$m_{t^\prime}>404$ GeV by ATLAS~\cite{Aad:2012xc}.
The mass splittings between up- and down-type fermions,
such as $m_{t'}-m_{b'}$ and $m_{\nu_{4}}-m_{e_{4}}$  are
constrained to be small by
the observables of the electroweak precision test,
such as the oblique parameters $S$ and $T$~\cite{Erler:2010sk}.

 In the SM4,  the fourth fermions have large Yukawa couplings to the Higgs boson,
 which leads to significant modifications to the predictions for
 the loop-induced processes such as gluon fusion $gg\to h$ and the decay $h\to \gamma\gamma$.
 Thus the  searches for Higgs boson production and decays can place stringent  constraints on the parameter space of  SM4, as the current experimental results are consistent with a SM-like Higgs boson within errors. The cross-section for Higgs boson production through $gg\to h$ is enhanced compared with that in the SM~\cite{Li:2010fu}.
One thus expects significant enhancements of the event rates of $gg\to h \to WW^{*}$, and $ZZ^{*}$, etc., which is not confirmed by the current data. On the other hand,
the partial decay width of $h\to \gamma\gamma$ is suppressed  in the SM4,
due to the destructive interference between the $W$-loop and $t'(b')$-loop.
Such a cancellation makes the prediction sensitive to
the next leading order (NLO) electroweak corrections
proportional to $G_{F} m_{t'(b')}^{2}$. The recent calculations show that  when the NLO electroweak corrections
are included, for heavy fourth quarks around $\sim 600$ GeV, the partial decay width $\Gamma(h\to \gamma\gamma)$ is  only $\sim 2-3\%$ of that in the SM~\cite{Denner:2011vt,Djouadi:2012ae,Lenz:2013iha}, which  makes the  event rate of $gg\to h \to \gamma\gamma$ far below the SM value and almost undetectable at LHC. 

In this work, we consider the SM4 in which
the fourth generation fermions and the SM fermions have different symmetry properties,
such that the transitions to the first three generation fermions are forbidden~\cite{ Erler:2010sk,Zhou:2011fr,Lee:2012xn,Murayama:2010xb}.
We show that in this scenario
the above mentioned tensions can be relaxed:
i)  the current direct search lower bounds  are no longer valid
as the fourth quarks cannot decay into the SM ones,
allowing  for relatively light fourth quarks.
ii) for  relatively light fourth generation quarks around 200 GeV the destructive interference between the fourth generation quark loops and the $W$-loop in the decay $h\to \gamma\gamma$ at NLO is reduced by an order of magnitude compared with $\sim600$ GeV fourth quarks, which relaxes the corresponding constraint.
iii)  due to the protection of the symmetry,
the lightest fourth neutrino can be a stable Majorana particle which
can be as light as 40 GeV without violating the LEP-II bounds.
Such a light stable neutrino can trigger
the Higgs invisible decay through $h\to \nu_{4}\bar{\nu}_{4 }$, which enhances the total width of the Higgs boson and
relaxes the constraints from the measurements of $gg\to h \to WW^{*}$, and $ZZ^{*}$, etc. Furthermore, the stable fourth neutrino can be a subdominant DM component which can be detected by the DM direct detection experiments. Through a  global $\chi^{2}$ analysis of the current LHC data on the Higgs production and decays,
 we obtain the allowed range of the  fourth Majorana neutrino mass and
 the mixing angle between the left-hand neutrino and right-hand anti-neutrino.
We find that the prediction for the recoil event rate is within the reach of the up coming direct detection experiments such as Xenon1T.

This paper is organized as follows.
 A brief introduction of SM4 with additional symmetries is given in section II. In section III, we study the phenomenology of the Higgs production and decays with focus on the invisible decay mode and compare the predictions with the experimental data of ATLAS and CMS through a global fit to the data. In Section IV, we study the stable fourth-generation Majorana neutrino as the dark matter candidate for its contribution to the DM relic density and predictions for the DM direct detection. Finally, a conclusion is given in section V.

\section{Fourth generation models with a stable Majorana neutrino}
We begin with a brief overview of the SM4.
In this type of model  the SM is extended with
an additional sequential fermion generation including a right-hand fourth neutrino.
\begin{equation}
\left(\begin{array}{c}t^\prime\\ b^\prime\end{array}\right)_L,\, t^\prime_R, \, b^\prime_R,\,  \left(\begin{array}{c}\nu_4\\ e_4\end{array}\right)_L,\, \nu_{4R}, \, e_{4R}.
\end{equation}
The fourth generation neutrinos can have both Dirac and Majorana mass terms. In the basis of $(\nu_L, \nu_R^c)^T$, the mass matrix for the fourth neutrino is given by
\begin{align}
m_\nu=\begin{pmatrix}
0 & m_D \\
m_D & m_M
\end{pmatrix} ,\label{eq:neutrinomass}
\end{align}
where $m_D$ and $m_M$ are the Dirac and Majorana masses, respectively. The left-hand components  $(\nu_{1L}^{(m)}, \nu_{2L}^{(m)})$ of the two mass eigenstates are related to the ones in the  flavor eigenstates by a rotation angle $\theta$
\begin{align}
\nu^{(m)}_{1L}& =-i (c_{\theta} \nu_L -s_{\theta} \nu_R^c) , \quad
\nu^{(m)}_{2L} =s_{\theta}\nu_L +c_{\theta} \nu_R^c ,
\end{align}
where $s_{\theta}\equiv\sin\theta$ and  $c_{\theta} \equiv\cos\theta$. The value of $\theta$ is defined in the range $(0,\pi/4)$ and  is determined by
\begin{align}
\tan 2\theta=\frac{2m_D}{m_M},
\end{align}
with $\theta=0 \ (\pi/4)$ corresponding to the limit of minimal (maximal) mixing.
The phase $-i$ in the expression of $\nu_{1L}^{(m)}$ is introduced to render the two mass eigenvalues real and positive.
The two Majorana mass eigenstates are $\chi_{1(2)}=\nu_{1(2)L}^{(m)}+\nu_{1(2)L}^{(m)c}$.
The masses of the two neutrinos are given by
$m_{1,2}=(\sqrt{m_M^2+4 m_D^2}\mp m_M)/2$. In terms of the mixing angle $\theta$, they
can be rewritten as
\begin{align}
m_1=\left(\frac{s_{\theta}}{c_{\theta}}\right) m_D, \quad
\mbox{and} \quad  m_2=\left(\frac{c_{\theta}}{s_{\theta}}\right) m_D ,
\end{align}
with $m_1\leq m_2$.  Note that for all the possible values of $\theta$ the lighter neutrino mass eigenstate $\chi_1$ consists of more left-hand component than the right-hand one, which means that $\chi_1$  always has sizable coupling to the SM $Z$-boson. Therefore the LEP-II  bound on the mass of stable neutrino is always valid for $\chi_1$, which is insensitive to  the mixing angle.

In the mass basis the interaction between the massive neutrinos and the SM $Z$-boson is given by
\begin{align}
 \mathcal{L}_{NC}= \frac{g_1}{4\cos\theta_W}
  \left[
    -c_{\theta}^2 \bar{\chi}_1\gamma^\mu\gamma^5 \chi_1-s_{\theta}^2 \bar{\chi}_2\gamma^\mu\gamma^5 \chi_2
    +2i c_{\theta}s_{\theta} \bar{\chi}_1 \gamma^\mu \chi_2
  \right]Z_\mu ,
\end{align}
where $g_1$ is the weak gauge coupling and $\theta_W$ is the Weinberg angle.
The Yukawa interaction between $\chi_{1,2}$ and the SM Higgs boson is given by
\begin{align}
\mathcal{L}_Y=-\frac{m_1}{v} \left(\frac{c_{\theta}}{s_{\theta}}\right)
\left[
c_{\theta}s_{\theta} \bar{\chi}_1\chi_1+c_{\theta}s_{\theta} \bar{\chi}_2\chi_2-i(c_{\theta}^2-s_{\theta}^2)\bar{\chi}_1 \gamma^5 \chi_2
\right] h\label{eq:neutrinoyukawa},
\end{align}
where $v\simeq246$ GeV is the vacuum expectation value.
 
The lighter fourth neutrino $\chi_{1}$ can be stable as a dark matter particle. This can be realized by introducing additional symmetries to the SM4, for instance:
\begin{itemize}
\item
An $ad hoc$ $Z_{2}$ symmetry under which
the fourth generation fermions are odd and the SM fermions are even. In some models, the $Z_2$ symmetry can be connected with the discrete $P$ and $CP$ symmetries of quantum fields~\cite{Guo:2010sy,Guo:2010vy,Guo:2008si}. Note that a discrete symmetry without gauge origin may be eventually broken by the effects of quantum gravity at Planck scale~\cite{Krauss:1988zc}.
\item
Another possibility is to introduce a new $U(1)$ gauge symmetry. All the fermions in SM4 are vector-like in the new gauge interaction. Through appropriately arranging the $U(1)$ charges of the fermions, the gauge anomaly can be canceled out among the generations~\cite{Zhou:2011fr,Zhou:2012dh}.

\item
In the SM, the local symmetry of $B-L$ which is the difference between
baryon number $B$ and lepton number $L$ and the hypercharge $Y$ are
know to be anomaly free for each generation. Thus one can assign a non-zero
$(B-L)+\alpha Y$ charge to the fourth generation fermions, where $\alpha$ is
a mixing parameter. For $\alpha=0$ case, the SM Higgs boson can give masses
to all the fermions. For $\alpha \neq 0$ case, additional Higgs boson has to be
introduced to generate the masses of the fourth fermions~\cite{Lee:2011jk,Lee:2012xn}.
\end{itemize}

In models with additional $U(1)$ gauge symmetry, new gauge boson $Z^\prime$ appears inevitably. The mass of $Z^\prime$ and its coupling to SM fermions as well as the mixing with $Z$ boson are constrained by various experiments (for a review, see~\cite{Langacker:2008yv}). In this work, we assume that the $Z^\prime$ boson is heavy enough, such that the current limits on $Z^\prime$ boson can be avoided, the fourth neutrino main interact with SM fermions though $Z$ and $h$.
%
\section{Phenomenology of Higgs production and decays}
%
According to the theory of QCD, the quarks and gluons are the fundamental degrees of freedom to participate in strong interactions at high energy. The QCD parton model plays a pivotal role in understanding hadron collisions. Due to the gluon luminosity, the gluon fusion is the main production channel for Higgs boson in proton-proton collisions throughout the entire Higgs mass range~\cite{Spira:1995rr}.
The leading order cross-section of $gg\to h$ at parton level is expressed as
\begin{eqnarray}
\hat{\sigma}_{gg\to h} & = & \frac{G_{F}\alpha_{s}^{2}}{288\sqrt{2}\pi}\left|\frac{3}{4}\sum_{q}A_{f}(\tau_{q})\right|^{2}\label{eq:gg2h}.
\end{eqnarray}
The fermion loop amplitude $A_f$ has the form
\begin{eqnarray}
&&A_{f} (\tau) =  2\left[\tau+(\tau-1)F(\tau)\right]/\tau^{2},\label{eq:af}\\
&&F (\tau)= \left\{ \begin{array}{ll}
\arcsin^{2}\sqrt{\tau} & \tau\leq1\\
-\frac{1}{4}\left[\log\frac{\sqrt{\tau}+\sqrt{\tau-1}}{\sqrt{\tau}-\sqrt{\tau-1}}-i\pi\right]^{2} & \tau>1
\end{array}\right. ,\label{eq:ft}
\end{eqnarray}
and the  scaling variable $\tau_i$ is defined as $\tau_i={m_h^2}/{(4m_i^2)}$.

The dependence of $A_f$ on the quark mass is rather weak when the fourth generation quarks are as heavy as few hundreds GeV. Thus it is unlikely to study the quark mass spectrum through Higgs boson searching. On the other hand, it is expected that the perturbative expansion breaks down for Yukawa couplings near or less than the perturbative unitarity bound, which allows maximal Dirac fermion masses of roughly $m_f \sim 600$ GeV~\cite{Denner:2011vt,Djouadi:2012ae,Bulava:2013ep,Lenz:2013iha}. For the fourth generation protected by the symmetry discussed in the previous section, the fourth generation quarks mass can be well below the bound safely, for instance $\sim200$ GeV. The production cross-section with NNLO QCD corrections has been implemented in the package HIGLU~\cite{Spira:1995mt}.

In addition to the enhancement of the production, the fourth generation fermions also change the total decay width and branch ratios of the Higgs boson. In order to compare the Higgs boson search signals in SM4 with the experimental results, the signal strength is defined as the cross-section of a given channel nomorlized to the SM expectation.
\begin{equation}
\mu_{i}^{\rm SM4}=\frac{\sigma^{\rm SM4}(pp\to h)}{\sigma^{\rm SM}(pp\to h)}\times\frac{\Gamma^{\rm SM}_{\rm tot}}{\Gamma^{\rm SM4}_{\rm tot}}\times\frac{\Gamma^{\rm SM4}_{i}}{\Gamma^{\rm SM}_{i}}.
\end{equation}
In this paper, we focus on the channels of $h\to WW^*/ZZ^*$ which are related to the test of electroweak symmetry breaking and the channel of $h\to\gamma\gamma$ which is the golden channel for light Higgs searching in SM. The experimental results $\mu_i^{\rm exp}$ are shown in Table.~\ref{tab:statsummary}. The two results of $h\to \gamma\gamma$ at CMS are based on different analysis, Multivariate (MVA) and Cut-based(CiC). The MVA approach gives about 15\% better expected sensitivity and the reusut is taken as the baseline result. 

\begin{table}[!t]
\begin{tabular}{|c|c|c|c|}
\hline
    & $h\to \gamma\gamma$ &$h\to WW^{*}$ & $h\to ZZ^*$\\
\hline
ATLAS& $1.57\pm0.22^{+0.24}_{-0.18}$ & $1.01\pm0.21\pm 0.19\pm0.13$ & $1.7^{+0.5}_{-0.4}$ \\
\hline
\multirow{2}{*}{CMS} & {$0.78^{+0.28}_{-0.26}$~(MVA)}  & \multirow{2}{*}{$0.76\pm0.21$} & \multirow{2}{*}{$0.91_{-0.24}^{+0.30}$}\\
&$1.11^{+0.32}_{-0.30}$~(CiC)& & \\
\hline
\end{tabular}
\caption{The Higgs searching results at the LHC in the channels $h\to\gamma\gamma$, $h\to WW^{(*)}$ and $h\to ZZ^{(*)}$ from ATLAS~\cite{G.Marchiori} and CMS~\cite{M.Pieri}. \label{tab:statsummary}}
\end{table}

\begin{figure}[htb]
\includegraphics[scale=1.5]{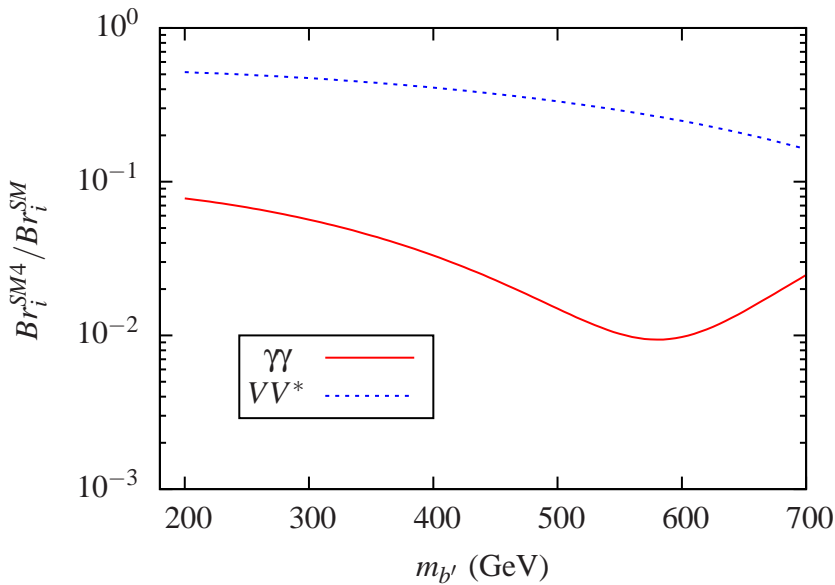}
\caption{The decay branch ratios of a 125 GeV Higgs boson into $\gamma\gamma$ and $V V^*$ states (with $V =W, Z$) as functions of $m_{b^\prime}$ in SM4 normalized to their SM values. }\label{fg:nlocorr}
\end{figure}

The Higgs decay to di-photon is mediated by $W$-boson and heavy charged fermions at one-loop level. The partial decay width can be written as
\begin{equation}
\Gamma(h\to2\gamma)=\frac{G_{F}\alpha^{2}m_{h}^{3}}{128\sqrt{2}\pi^{3}}\left|\sum_{f}N_{C}Q_{f}^{2}A_{f}(\tau_{f})+A_{W}(\tau_{W})\right|^{2},
\end{equation}
where the $f$ in the summation denotes all the massive charged fermions including quarks and leptons. The $A_{f}$ as shown in Eq.~(\ref{eq:af}) is coming from the fermion loop while the $A_{W}$ coming from the $W$ loop expressed as follows
\begin{eqnarray}
A_{W} & = & -\left[2\tau_W^{2}+3\tau_W+3(2\tau_W-1)F(\tau_W)\right]/\tau_W^{2}.
\end{eqnarray}
With a 125 GeV Higgs boson which is below the $WW$ and charged fourth family fermion pair threshold, the $A_{W}$ and $A_{f}$ are real but with opposite signs. Due to the destructive interference, the partial width in SM4 is much smaller than that in SM. The recent calculations show that  when the NLO electroweak corrections
are included, for heavy fourth quarks around $\sim 500$ GeV, the partial decay width $\Gamma(h\to \gamma\gamma)$ is  only $\sim 2-3\%$ of that in the SM~\cite{Denner:2011vt,Djouadi:2012ae,Lenz:2013iha}, which  make the  event rate of $gg\to h \to \gamma\gamma$ far below the SM value. 
However, the NLO electroweak corrections strongly depend on the mass of the fourth generation fermions. In Fig.~\ref{fg:nlocorr}, we show ${\rm Br}(h\to\gamma\gamma)$ for different values of $m_{b^\prime}$ with the mass splitting $m_{t^\prime}-m_{b^\prime}\simeq50 \GeV$~\cite{Kribs:2007nz}. In the calculations, the numerical package HDECAY~\cite{Djouadi:1997yw} which includes NLO electroweak corrections is used.
One can see that at the case of $m_{b^\prime}\sim 600 \GeV$, the branch ratio of $h\to\gamma\gamma$ is only $\sim1\%$ of the SM value. But at $m_{b^\prime}\sim200$ GeV, the branch ration in SM4 is close to $10\%$. Since the production cross-section can be enhanced by a factor of about $10$ including NNLO QCD corrections, the final signal strength in SM4 can still be comparable with that in the SM.

\begin{table}[htb]
\begin{tabular}{c|c|c|c|c|c|c}
\hline
& $b\bar{b}$ & $gg$ & $Z\gamma$ & $\gamma\gamma$ & $WW^*$ & $ZZ^*$ \\ 
\hline
SM & 2.35 &0.349 & $6.26\times 10^{-3}$ &$9.27\times10^{-3}$ & 0.882 & $10.8\times10^{-2}$ \\
\hline
SM4 & 2.46 &3.55 & $4.89\times10^{-3}$ &$1.38\times10^{-3}$ & 0.828 & $9.92\times 10^{-2}$\\
\hline
\end{tabular}
\caption{The partial widths in units of MeV of the Higgs boson decay($m_h=125$ GeV) in SM and SM4 obtained with HDECAY~\cite{Djouadi:1997yw}.\label{tb:par_width}}
\end{table}

Another important channel modified in SM4 is $h\to gg$. 
The process is mediated by heavy quarks in loop, where the main contribution is coming from top quark
and small contribution from bottom quark in SM. In SM4, the fourth generation quarks would enhance this channel to a level where it can be competitive with the $b\bar{b}$ decay mode.
The channel $h\to Z\gamma$ is also modified at leading order by the fourth generation fermions.
But the contribution is in general small. The channels $h\to ZZ^{*}(WW^*)$ and $h\to b\bar{b}(s\bar{s},c\bar{c},\tau^+\tau^-,\mu^+\mu^-)$ in SM4 have the same partial widths as in SM at leading order, and can be calculated using HDECAY with NLO electroweak corrections.

\begin{figure}[htp]
\includegraphics[width=0.48\textwidth]{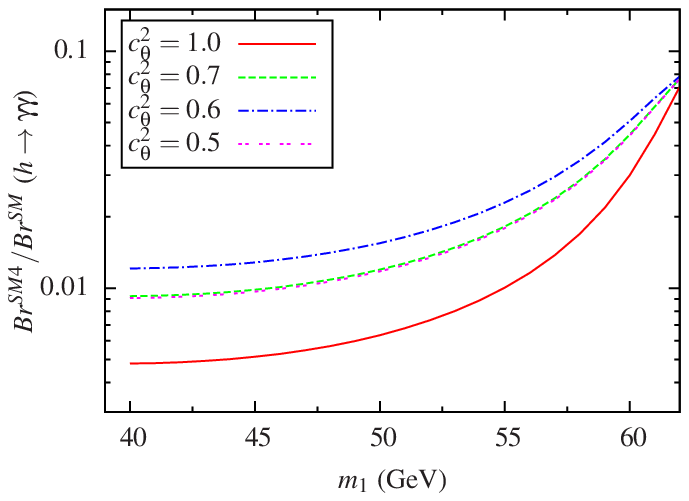}
\includegraphics[width=0.48\textwidth]{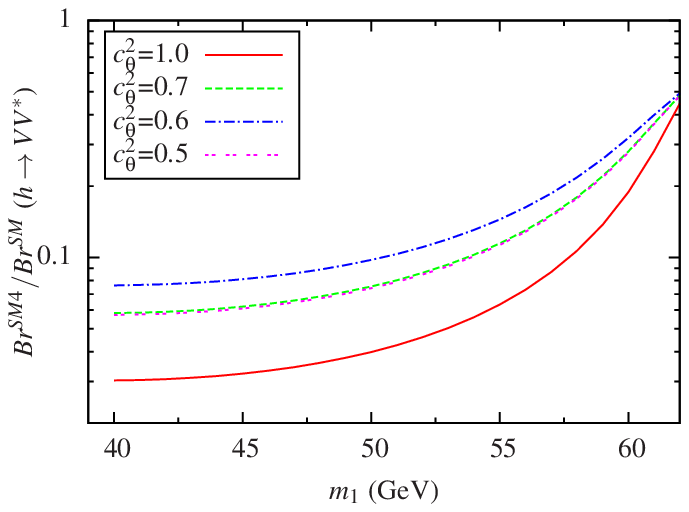}
\caption{The branching ratios of $h\to\gamma\gamma$ and $h\to VV^{*}$ (with $V=W,Z$) as functions of $m_1$ and $c^2_\theta$ in SM4 normalized to their SM values. }\label{fg:branch}
\end{figure}

The fermions in SM obtain masses though Yukawa interaction with the Higgs boson. If the neutrino in fourth generation is Dirac particle as other fermions, its Yukawa coupling should be $m_\chi/v$. However, the neutrino could be Majorana particle. In this case, the Yukawa coupling is only a part of the mass term as shown in Eq.~(\ref{eq:neutrinomass}). The partial widths of the Higgs boson decay to Majorana neutrinos are given by
\begin{eqnarray}
\Gamma(h\to \chi_1\chi_1)&=&\frac{1}{2}\frac{(2c^2_\theta)^2 G_{F}m_{h}}{4\sqrt{2}\pi}m_1^{2}\left(1-4m_1^{2}/m_{h}^{2}\right)^{3/2},\nonumber\\
\Gamma(h\to\chi_2\chi_2)&=&\frac{1}{2}\frac{(2c^2_\theta)^2 G_{F}m_{h}}{4\sqrt{2}\pi}m_1^{2}\left(1-4m_2^{2}/m_{h}^{2}\right)^{3/2},\nonumber\\
\Gamma(h\to\chi_1\chi_2)&=&\frac{G_{F}m_{h}}{4\sqrt{2}\pi}\frac{c^2_\theta(c^2_\theta-s_\theta^2)^2}{s^2_\theta}m_1^{2}\left(1-\frac{(m_1+m_2)^{2}}{m_{h}^{2}}\right)^{3/2}\left(1-\frac{(m_1-m_2)^{2}}{m_{h}^{2}}\right)^{1/2}.\label{eq:h2ff}
\end{eqnarray}
The partial decay widths exhibits a strong suppression near the thresholds. If the neutrino is Dirac fermion ($\theta=\pi/4$) and its mass is $50~\GeV$, the partial width is about $44$ MeV which would be the dominated contribution to the total width of the Higgs boson. The branch ratios of $h\to\gamma\gamma$ and $h\to VV^* (V=W, Z)$ as functions of $c^2_\theta$ and $m_1$ are shown in Fig.~\ref{fg:branch}. 

\begin{figure}[hbt]
\includegraphics[scale=1.5]{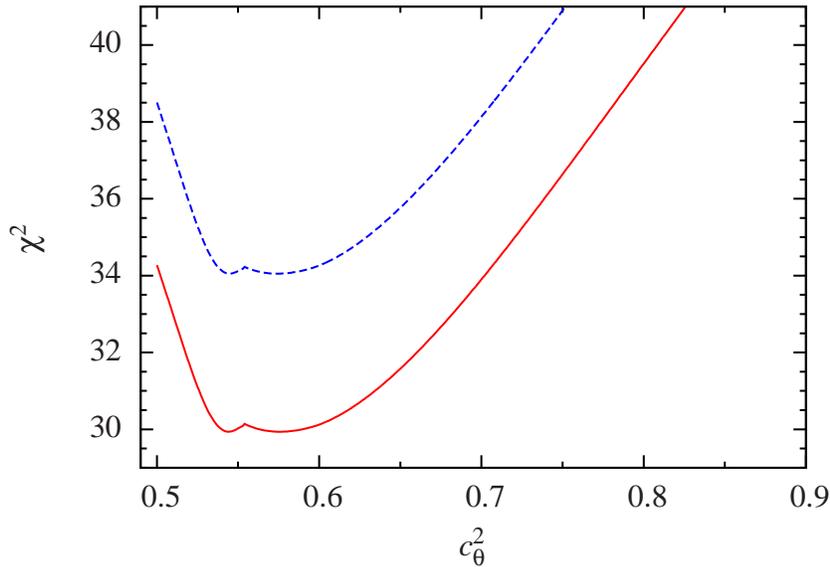}
\caption{The $\chi^2$ as a function of $c^2_\theta$ ($m_1=50$ GeV). The experimental value for the $h\to\gamma\gamma$ from CMS is chosen as $0.78^{+0.28}_{-0.26}$($1.11^{+0.32}_{-0.30}$)~\cite{M.Pieri} for the solid (dashed) line.}\label{fg:chis}
\end{figure}

We compare the signal strength $\mu_i$ in SM4 including invisible decay mode with the LHC data through a $\chi^2$ analysis. The $\chi^2$ is defined as
\begin{equation}
\chi^2\equiv\sum_{i}\left(\frac{(\mu^{\rm SM4}_i-\mu^{\rm exp}_i)^2}{(\sigma^{\rm exp}_i)^2}\right),
\end{equation}
which depends on the parameter $c^2_\theta$ and the mass of $\chi_1$. For a fixed $m_1=50$ GeV, the results of $\chi^2$ as a function of $c^2_\theta$ are shown in Fig.~\ref{fg:chis}. One can see there are two minimums for each line. For example, the two minimums of the solid line are obtained at $c_\theta^2=0.54$ ($m_2=59$ GeV) and $c_\theta^2=0.58$ ($m_2=68$ GeV).  At the case of $c_\theta^2=0.58$, the Higgs boson decay mode $h\to\chi_2\chi_2$ is not allowed.

\begin{figure}[hbt]
\includegraphics[scale=1.5]{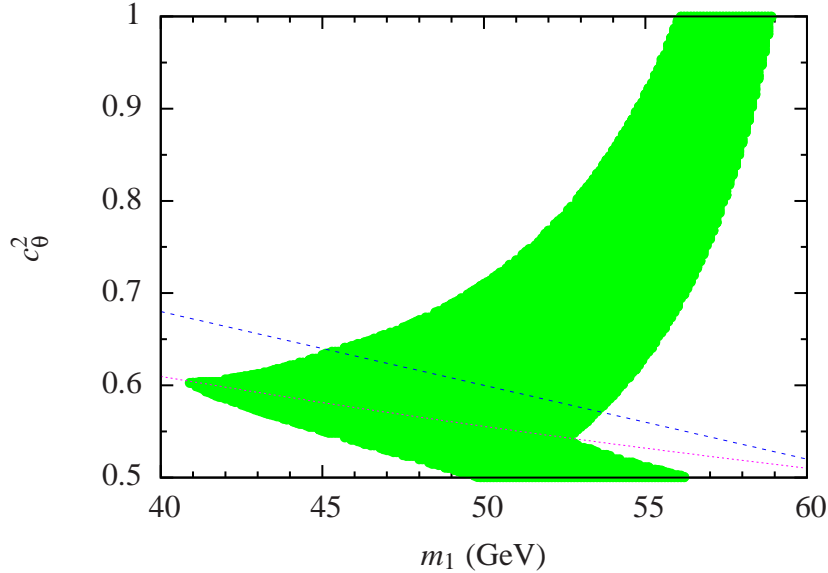}
\caption{The contour plot in the plane of $c^2_\theta$ and $m_1$ with 95\% C.L.. The dashed (dotted) line is the threshold for the channel $h\to\chi_1\chi_2$ ($\chi_2\chi_2$)}\label{fg:contour}
\end{figure}

The contour of allowed region in the plane of $m_1$ and $c^2_\theta$ is shown in Fig.~\ref{fg:contour}.  One can find allowed range $41<m_1<59$ GeV for the light Majorana neutrino mass at 95\% C.L.. The dashed (dotted) line is the threshold for the channel $h\to\chi_1\chi_2$ ($\chi_2\chi_2$) to be opened. Below the threshold, the decay mode $h\to\chi_1\chi_2$ ($h\to\chi_2\chi_2$) is not allowed.

\begin{figure}
\includegraphics[width=0.3\textwidth]{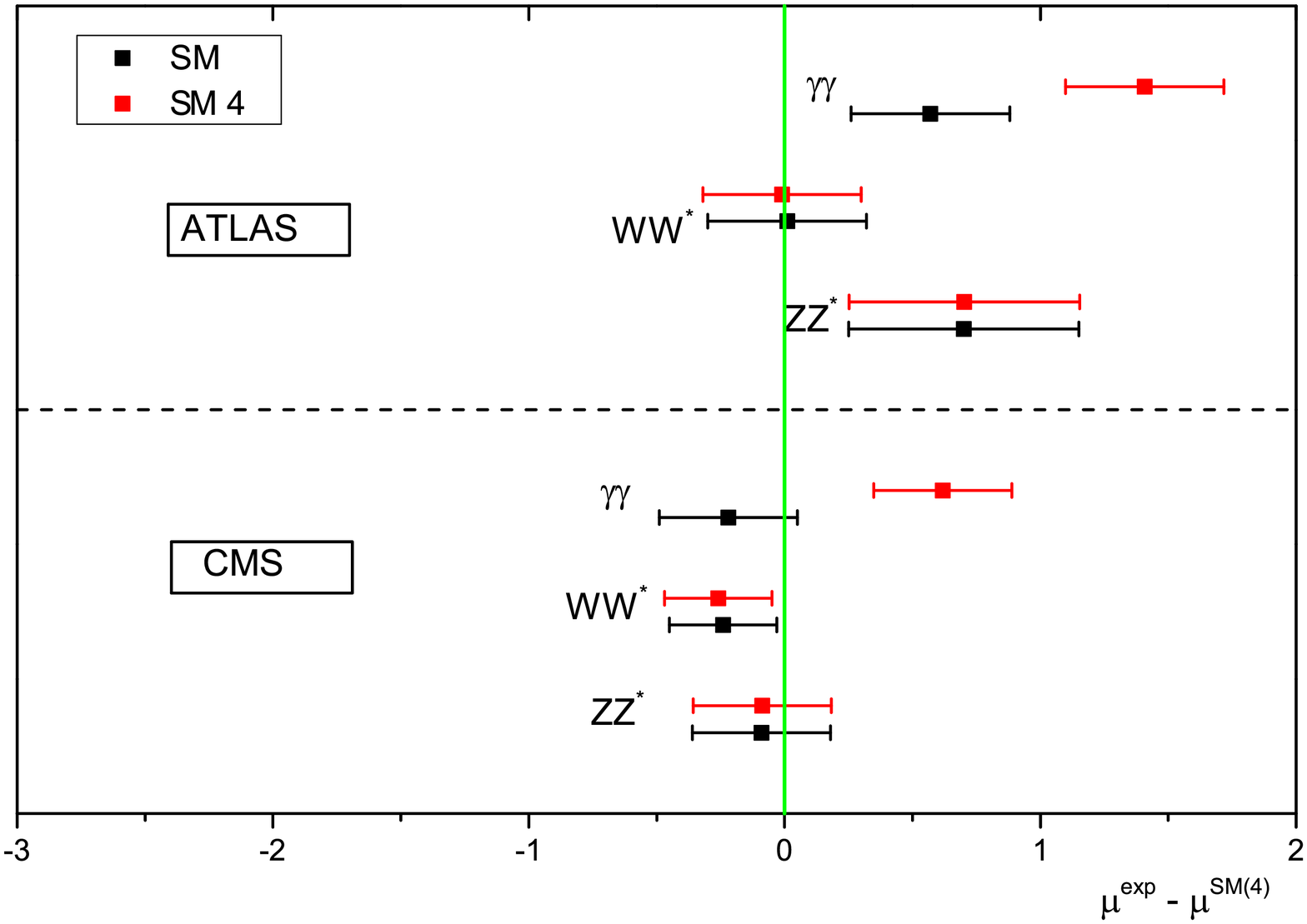}
\includegraphics[width=0.3\textwidth]{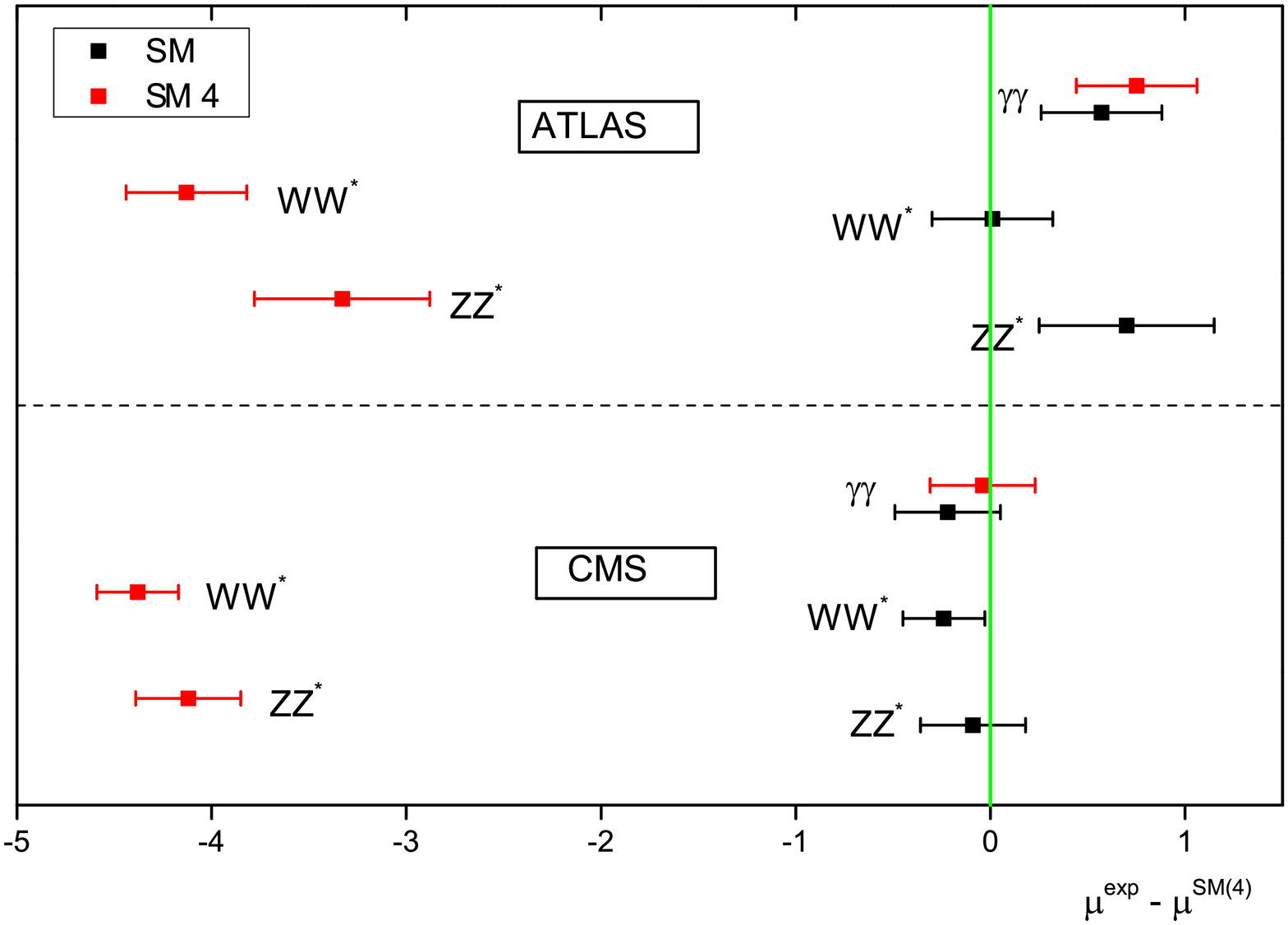}
\includegraphics[width=0.3\textwidth]{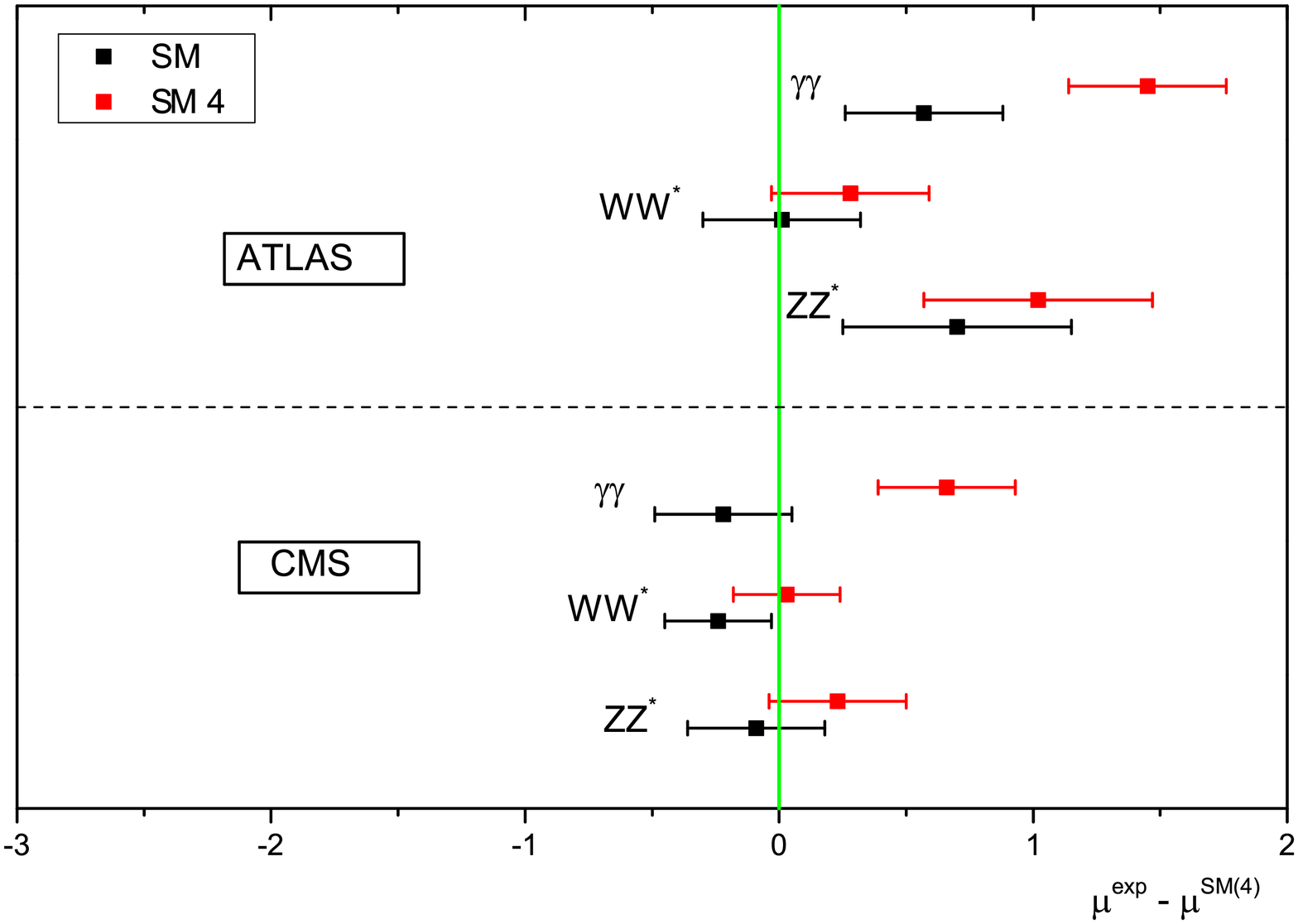}
\caption{The signal strength difference $\mu_i^{\rm exp}-\mu_i^{\rm SM(4)}$. (Left) for the case of minimal $\chi^2$, (Middle) for the case that the invisible decay mode of the Higgs boson is not considered, and (Right) for the case of $c_\theta^2=0.5$ (the fourth neutrino is Dirac type fermion).}\label{fig:biassm4}
\end{figure}

In Fig.~\ref{fig:biassm4}, we show the strength difference $\mu^{\rm exp}_i-\mu^{\rm SM(4)}_i$ between the theoretical prediction and experimental values.
 One can see that without invisible decay (Middle in Fig.~\ref{fig:biassm4}), the SM4 predictions deviate from the experimental values of ATLAS and CMS largely, especially for the $h\to ZZ^*(WW^*)$. When the invisible decay is considered, at the case of best fit (Left in Fig.~\ref{fig:biassm4}), one can see that the prediction is consistent with the experimental observations within $2\sigma$, except the $h\to \gamma\gamma$. Since the current data of ATLAS and CMS are not fully consistent with each other, future more accurate LHC measurement are useful in testing the SM4.

\section{Relic density and direct detection of the fourth generation neutrino dark matter}\label{relic-density}

Stable neutrinos heavier than $\sim 1$ GeV are
possible candidates for the cold DM~\cite{Lee:1977ua,Kolb:1985nn}.
However, it is well-known that for a neutrino heavier than 40 GeV,
the cross-section for its annihilation into $f\bar f$, $W^{\pm}W^{\mp}$, $ZZ$ or $Z h$
is in general too large to reproduce the observed DM relic density~\cite{Enqvist:1988we}.
Thus the fourth neutrino can only be a subdominant component of the
whole DM in the Universe. From theoretical point of view,
it is natural to have multi-component DM, as the lightest SM neutrino is already known to be
a subdominant  dark matter component.
In some multi-component DM models,
the interactions between the DM components actually can
provide  a new source of boost factor required to
explain the electron/positron excesses observed by
the experiments like PAMELA, Fermi-LAT, HESS and AMS-02 \cite{Liu:2011aa,Liu:2011mn,Jin:2013nta}.
Despite its very low number density,
the fourth neutrino can still be probed by
the  DM direct detection experiments
due to its relatively strong gauge (Yukawa) coupling to the target nuclei through $Z$ ($h$) exchange,
which provides an alternative way to search for new physics
beyond the SM complementary to that at the LHC.
For a subdominant dark matter particle,
the event rate of the DM-nucleus elastic scattering depends on
the its fraction in the halo DM density and the cross-section of the scattering. Both of them have nontrivial dependences on the  neutrino mass
and the mixing angle $\theta$~\cite{Zhou:2011fr,Zhou:2012dh}.

The thermal relic density of $\chi_1$ is related to its annihilation cross-section at freeze out.
When $\chi_1$ is lighter than the $W^\pm$ boson,
$\chi_1\chi_1$ can only annihilate into
light SM fermion pairs $f\bar{f}$ ($f=u, d, c, s, b$) through $s$-channel $Z/h$ exchange.
For Majorana neutrino the annihilation cross-sections are suppressed by
the masses of the final state fermions.
However,
large enhancement of the annihilation cross-section occurs
when the mass of $\chi_1$ is close to $m_Z/2$($m_{h}/2$)
such that the intermediate state $Z$ ($h$) is nearly on shell.
In order to determine the DM relic density,
one needs to calculate the thermally averaged product of
the DM annihilation cross-section and the relative velocity
which is given by
\begin{align}
\langle \sigma v \rangle
=\frac{1}{8m_1^2 T K^2_2(m_1/T)}\int_{4m_1^2}^{\infty} ds \sigma (s-4m_1^2) \sqrt{s} K_1
\left(\frac{\sqrt{s}}{T} \right) ,
\end{align}
where  $T$ is temperature of the thermal bath and
$K_{1,2}(x)$ are the modified Bessel function of the second kind.
The relic abundance can be approximately estimated as
\begin{align}
\Omega h^2 \simeq \frac{1.07\times 10^9 \mbox{GeV}^{-1}}{\sqrt{g_*} M_{\text{pl}} \int_{x_F}^{\infty} \frac{\langle \sigma v \rangle}{x^2}dx} ,
\end{align}
where $x=m_1/T$ is the rescaled inverse temperature,
and $x_F\approx 25$ corresponds to the decoupling temperature.
The number of effective relativistic degrees of freedom
at the time of  freeze out is  $g_*=86.25$,
and $M_{\text{pl}}=1.22\times 10^{19}$ GeV is the Planck mass scale.

\begin{figure}[htb]
\begin{center}
\includegraphics[width=0.45\textwidth]{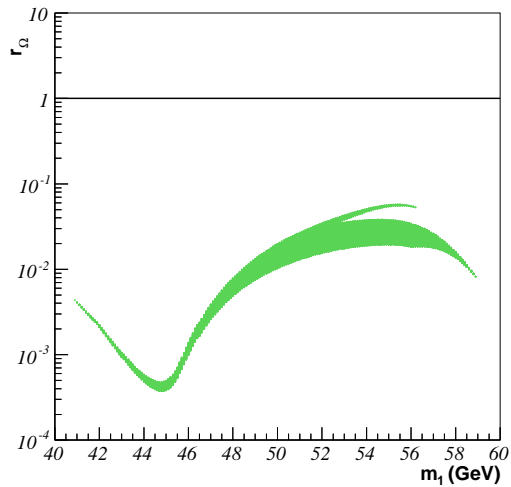}
\caption{
  The rescaled $\chi_1$ relic density $r_\Omega$  as a function of the mass of $\chi_1$.}
\label{fig:cross-section}
\end{center}
\end{figure}

We calculate the cross-sections for
$\chi_1\chi_1$ annihilation into all the relevant final states and
the thermal relic density
using the numerical package micrOmegas 2.4~\cite{Belanger:2010gh}.
In Fig.~\ref{fig:cross-section}
we show the quantity $r_\Omega\equiv \Omega_{\chi_{1}}/\Omega_{\rm DM}$
which is the ratio of
the relic density of $\chi_1$ to the observed total DM relic density
$\Omega_{\rm DM} h^2=0.110\pm0.006$~\cite{Beringer:1900zz}
for the allowed values of $m_{1}$ and $\theta$ determined from
the previous section.
The result  clearly shows that due to the large annihilation cross-section,
$\chi_1$ cannot make up the whole DM in the Universe.
Its contribution is  less than  $10\%$ of the total DM relic density.
However,
since $\chi_1$ has relatively strong couplings to $h$ and $Z$,
even in the case that the number density of $\chi_1$ is very low in the DM halo, it is still possible that it can be detected by
its elastic scattering off nucleus in direct detection experiments.
Given the difficulties in detecting such a neutral and stable particle at the LHC,
it is possible that the stable fourth generation neutrino could be
first seen  or ruled out at the DM direct detection experiments.

The differential event rate of DM-nucleus scattering per nucleus mass is given by
\begin{align}\label{eq:event-rate}
\frac{dN}{dE_R}=\frac{\rho_{\rm DM} \sigma_N}{2m_{\rm DM} \mu_N^2}F^2(E_{R})\int^{v_{\rm esc}}_{v_{\rm min}}d^3 v
\frac{f(v)}{v} ,
\end{align}
where $E_R$ is the recoil energy,
$\sigma_N$ is the scattering cross-section corresponding to the zero momentum transfer, $m_{\rm DM}$ is the mass of the DM particle,
$\mu_N=m_{\rm DM}m_{N}/(m_{\rm DM}+m_N)$ is the DM-nucleus reduced mass,
$F(E_{R})$ is the form factor,
and $f(v)$ is the velocity distribution function of the halo DM.
The local total DM density $\rho_{\rm DM}$ is often set to be equal to $\rho_0\simeq 0.3 \mbox{ GeV}/\mbox{cm}^3$ ( for updated determinations of $\rho_0$, see e.g. ~\cite{Catena:2009mf})
which is  commonly adopted by the current DM direct detection  experiments
as a benchmark value.
We assume that there is no difference in the clustering of DM
for the subdominant and dominant DM components
such that $\rho_1$ is proportional to the relic density of $\chi_1$ in
the Universe, namely
\begin{align}r_\rho\equiv \frac{\rho_1}{\rho_{0}}\approx
r_{\Omega}.
\end{align}
If the DM particles are nearly collisionless and
there is no long-range interactions acting differently on different DM components,
it is expected that the structure formation process should not
change the relative abundances of the  DM components.
Thus the expected event rate of the DM-nucleus elastic scattering will be simply scaled down by the factor $r_\rho$.
In order to directly compare the theoretical predictions with
the reported  experimental upper limits which
are often obtained under the assumption of $r_{\rho}=1$,
we shall calculate the rescaled elastic scattering cross-section
\begin{align}
\tilde{\sigma} \equiv   r_\rho \sigma \approx r_\Omega \sigma ,
\end{align}
which corresponds to the event rate  measured by the direct detection experiments.
Note that
$\tilde{\sigma}$ may depend on $m_{1}$ through the ratio $r_\rho$
even when the cross-section $\sigma$ is independent of the $m_1$.

The spin-independent DM-nucleon elastic scattering
cross-section in the limit of zero momentum transfer is given by~\cite{Jungman:1995df}
\begin{align}
\sigma^{SI}_n=\frac{4 \mu_n^2}{\pi }\frac{[Z f_p+(A-Z) f_n]^2}{A^2} ,
\end{align}
where $Z$ and $A-Z$ are the number of protons and neutrons within the target
nucleus, respectively. $\mu_n=m_1 m_n/(m_1+m_n)$ is the DM-nucleon reduced mass. The interaction between DM particle and the proton (neutron) is given by
\begin{align}
f_{p(n)}&=\sum_{q=u,d,s}f^{p(n)}_{Tq} a_q \frac{m_{p(n)}}{m_q}
+\frac{2}{27} f^{p(n)}_{TG}\sum_{q=c,b,t}a_q \frac{m_{p(n)}}{m_q} ,
\end{align}
with $f^{p(n)}_{Tq}$ the DM coupling to light quarks and
$f^{p(n)}_{TG}=1-\sum_{q=u,d,s}f^{p(n)}_{Tq}$.
In the case where
the elastic scattering is dominated by $t$-channel Higgs boson exchange,
the isospin conservation relation $f_n\simeq f_p$ holds and
one has $\sigma^{SI}_n \simeq 4 f_n^2\mu_n^2/\pi$.
In numerical calculations
we take $f^p_{Tu}=0.020\pm0.004$,
$f^p_{Td}=0.026\pm0.005$, $f^p_{Ts}=0.118\pm0.062$, $f^n_{Tu}=0.014\pm0.003$,
$f^n_{Td}=0.036\pm0.008$ and $f^n_{Ts}=0.118\pm0.062$~\cite{hep-ph/0001005}.
The coefficient $a_q$ in the model is given by
\begin{align}
a_q=c_{\theta}^2\frac{m_1 m_q}{v^2 m_h^2} .
\end{align}
The value of $a_q$ is proportional to $m_1$, thus larger elastic scattering
cross-section is expected for heavier $\chi_1$.  Note that in terms of $m_1$
the coefficient $a_q$ is proportional to $c_{\theta}^2$.  Part of the mixing
effects has been absorbed into the mass of $\chi_1$. In the limit of $\theta
\to 0$, $m_1$ is approaching zero and the coupling between $\chi_1$ and $h$
is vanishing as expected. The value of $a_q$ has a strong dependence on
$m_h$.
The quark mass $m_q$ in the expression of
$a_q$ cancels the one in the expression of $f_{p(n)}$. Therefore there is no quark mass dependence
in the calculations.

\begin{figure}[htb]
\begin{center}
\includegraphics[scale=0.4]{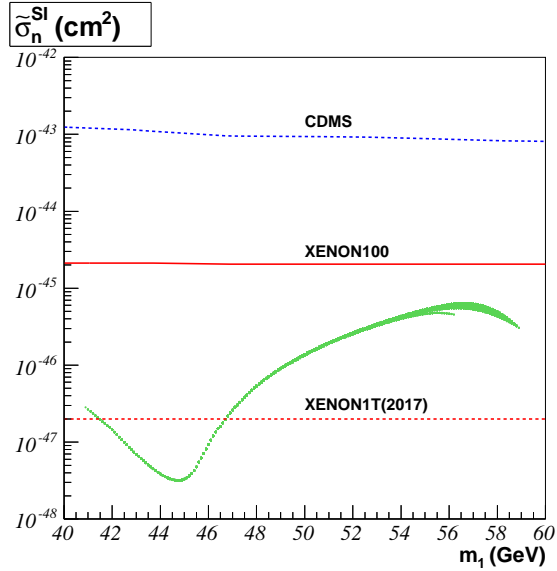}
\caption{Effective spin-independent cross-section $\tilde{\sigma}^{SI}_n$
  which is $\sigma^{SI}_n$ rescaled by $r_\rho\approx r_\Omega$ for $\chi_1$
  elastically scattering off nucleon as function of the mass of $\chi_1$.
  The current upper limits from CDMS~\cite{Ahmed:2009zw} and
  Xenon100~\cite{Aprile:2011hi} experiments are also shown.
}\label{fig:SIneutron}
\end{center}
\end{figure}

Using the allowed range of $m_{1}$ and $c^2_\theta$,
the predicted spin-independent effective cross-sections $\tilde{\sigma}^{SI}_n$ are obtained and shown in  Fig.~\ref{fig:SIneutron}.
For a given $m_{1}$,
the predicted $\tilde{\sigma}^{SI}_n$ is  found to be in a very narrow range,
which is due to a neary complete cancellation in the $\theta$-dependence between ${\rho}_0$
and the cross-section $\sigma^{SI}_{n}$,
as in the allowed range of $m_1$,
both the DM annihilation and DM-nucleus elastic scattering cross-section are
proportional to  $c_{\theta}^{4}$.
The insensitivity to the mixing angle leads to unambiguous prediction for $\tilde{\sigma}^{\rm SI}_n$. As can be seen from the figure,
the predicted effective  cross-section can be
close to the current Xenon100 upper bound for
$m_{1}\sim 56$ GeV.
The future Xenon1T experiment can probe most of
the allowed mass range of the fourth neutrino from 47 to 59 GeV.

The Majorana neutrino DM can contribute to spin-dependent elastic scattering cross-section through axial-vector current interaction induced by the exchange of the
$Z$ boson.
At the limit of the  zero momentum transfer, the spin-dependent cross-section has
the following form~\cite{Jungman:1995df}
\begin{align}\sigma_N^{SD}=\frac{32}{\pi}G_F^2 \mu_n^2 \frac{J+1}{J}
\left(a_p \langle S_p\rangle + a_n \langle S_n\rangle \right)^2 ,
\end{align}
where $J$ is the spin of the nucleus, $a_{p(n)}$ is the DM effective coupling to
proton (neutron) and $\langle S_{p(n)}\rangle$ the expectation value of the
spin content of the nucleon within the nucleus. $G_F$ is the Fermi constant.
The coupling $a_{p(n)}$ can be
written as
\begin{align}a_{p(n)}=\sum_{u,d,s}\frac{d_q}{\sqrt{2} G_F} \Delta^{p(n)}_q ,
\end{align}
where $d_q$ is the DM coupling to quark and $\Delta^{p(n)}_q$ is the fraction of the
proton (neutron) spin carried by a given quark $q$.
The spin-dependent DM-nucleon elastic scattering cross section is given by
\begin{align}
\sigma_{p(n)}^{SD}=\frac{24}{\pi} G_F^2 \mu_n^2
\left(  \frac{d_u}{\sqrt{2} G_F }\Delta^{p(n)}_u
+\frac{d_d}{\sqrt{2} G_F }\Delta^{p(n)}_d
+\frac{d_s}{\sqrt{2} G_F }\Delta^{p(n)}_s
\right)^2 .
\end{align}
In numerical calculations we take
$\Delta^p_u=0.77$, $\Delta^p_d=-0.40$, $\Delta^p_s=-0.12$~\cite{Cohen:2010gj},
and use the  relations $\Delta^n_u=\Delta^p_d$, $\Delta^n_d=\Delta^p_u$,
$\Delta^n_s=\Delta^p_s$. The coefficients $d_q$ in this model are given by
\begin{align}d_u=-d_d=-d_s=\frac{G_F}{\sqrt{2}} .
\end{align}
Note that
for the axial-vector current interactions,
the coupling strengths do not depend on the electromagnetic charges of the quarks.

\begin{figure}[htb]
\begin{center}
\includegraphics[width=0.45\textwidth]{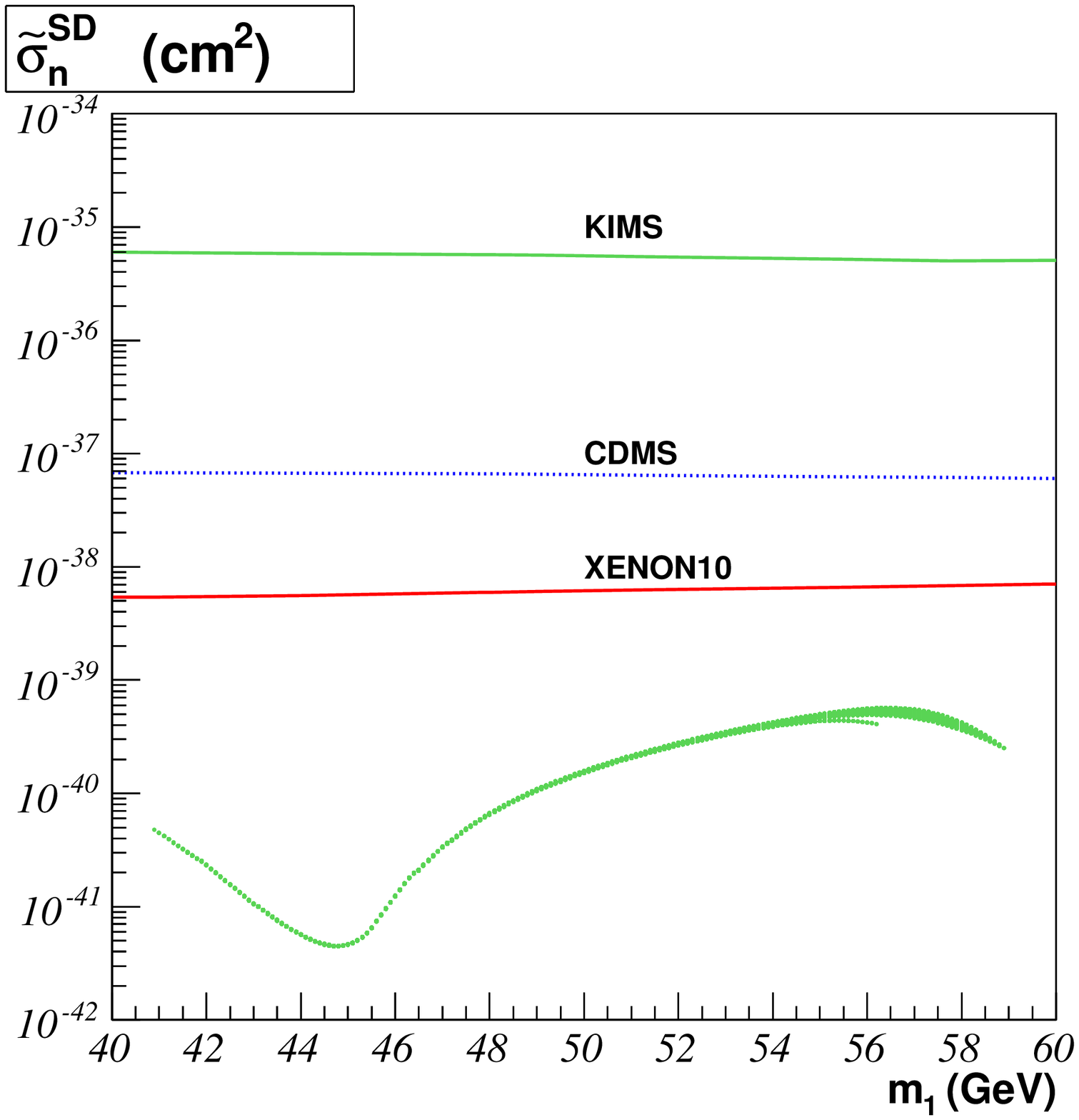}
\includegraphics[width=0.45\textwidth]{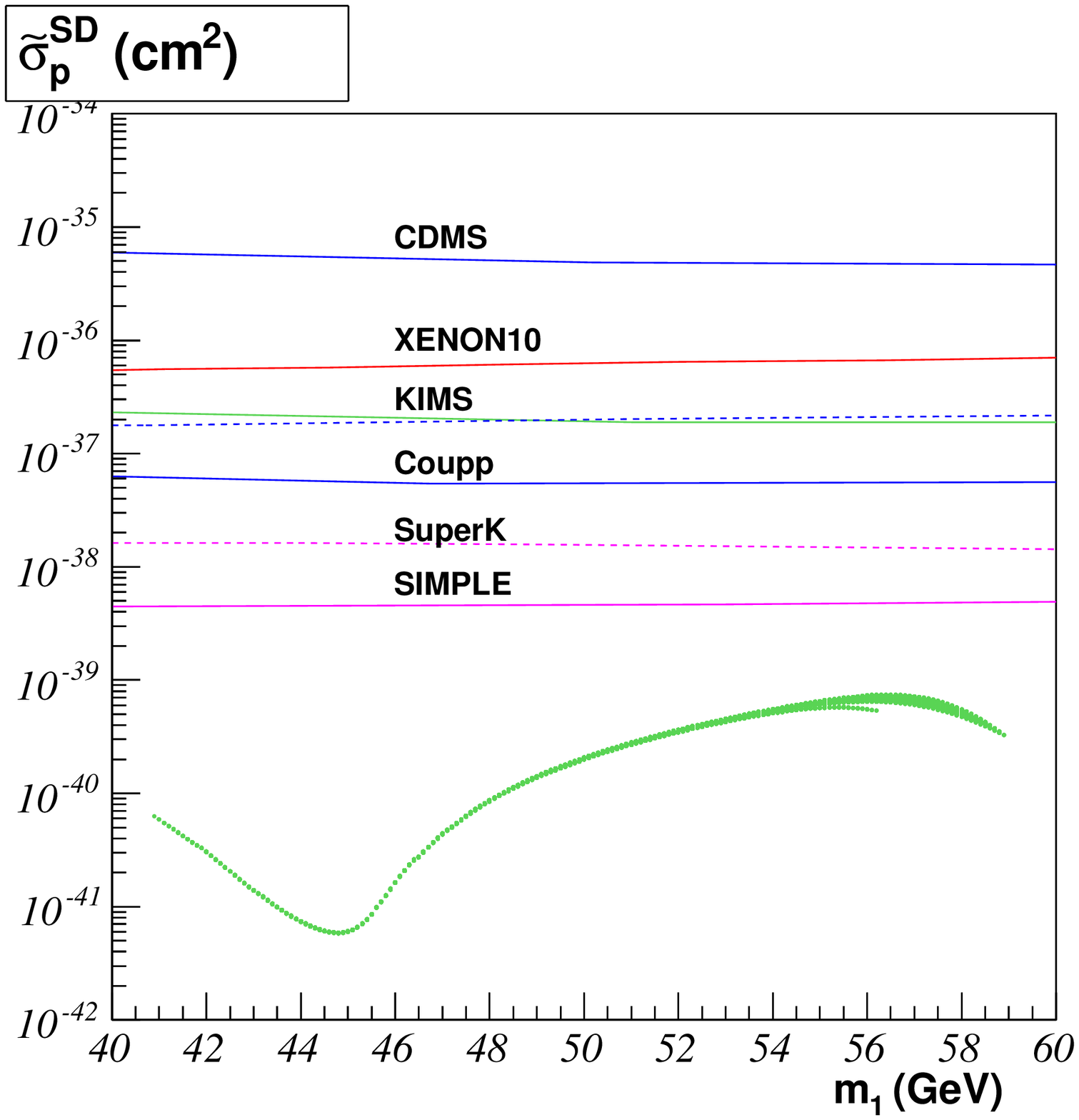}
\caption{
The effective spin-dependent cross-section $\tilde{\sigma}^{SD}_n$ (Left) and  $\tilde{\sigma}^{SD}_p$ (Right). The current upper limits from various experiments such as Xenon10~\cite{Angle:2008we}, KIMS~\cite{Lee.:2007qn}, CDMS~\cite{Akerib:2005za},  Coupp~\cite{Behnke:2010xt}, Picasso~\cite{Archambault:2009sm}, SIMPLE~\cite{Felizardo:2011uw}, SuperK~\cite{Desai:2004pq} and IceCube~\cite{0902.2460} are also shown.
 }
\label{fig:SDn}
\end{center}
\end{figure}

In  Fig.~\ref{fig:SDn}
we show the predicted effective spin-dependent DM-neutron (proton) cross-section
$\tilde{\sigma}_{n(p)}^{SD}$
together with various experimental upper limits.
The cross-sections for Majorana neutrino DM scattering off
proton and neutron are quite similar, which is due to the fact that the relative
opposite signs in $\Delta_u$ and $\Delta_d$ are compensated by the opposite
signs in $d_u$ and $d_n$. So far the most stringent limit on the DM-proton
spin-dependent cross-section is reported by the SIMPLE
experiment~\cite{1106.3014}.
Note that
  Different assumptions on the value of $r_\rho$  and the nature of the heavy stable neutrino may result in
  different limits.  For instance, in
  Ref.~\cite{Angle:2008we}, an excluded mass range of 10 GeV-2 TeV
  is obtained from the Xenon10 data on the cross-section of the
  spin-dependent DM-nucleus elastic scattering, which is based on the
  assumption that the local halo DM is entirely composed of stable Majorana
  neutrino, i.e. $r_\rho=1$, and the neutrino has the same couplings to the $Z$ boson as that
  of the SM active neutrinos. As in the present model we have $r_\rho \approx r_\Omega \ll 1$ and the coupling to the
$Z$ boson depends on the mixing angle, the resulting constraints are different significantly.

\section{Conclusions}\label{conclusion}
In summary, we have considered a type of extensions of the simplest SM4, in which extra symmetries are introduced to prevent the transitions of the fourth generation fermions to the ones in the first three generations.
In these models, the  lower bounds on
the masses of fourth generation quarks from direct searches can be relaxed, and at the same time the fourth neutrino is allowed to be stable and light enough to trigger the Higgs boson invisible decay. In addition, the fourth Majorana neutrino becomes a detectable dark matter particle.
We have performed a global analysis of the current LHC data on the Higgs boson production and decays in this type of SM4. The results show that the mass of the fourth Majorana neutrino is confined in the range $\sim 41-59$ GeV at 95\% C.L.. Within the allowed parameter space,  we have found that the predicted effective cross-section for spin-independent DM-nucleus scattering is
$\sim 3\times 10^{-48}-6\times 10^{-46} \text{ cm}^{2}$,
which is  insensitive to the mixing angle between
the left- and right-hand component of the Majorana neutrino,
and  can be tested by the Xenon1T experiment in the near future. The predicted spin-dependent cross-sections can reach $\sim 8\times 10^{-40}\text{ cm}^{2}$.

\begin{acknowledgments}
This work  is supported in part by
the National Basic Research Program of China (973 Program) under Grants No. 2010CB833000,
the National Nature Science Foundation of China (NSFC) under Grants No. 10975170, 10821504, 11275114,
the Project of Knowledge Innovation Program (PKIP) of the Chinese Academy of Science,
the Natural Science Foundation of Shandong Province,
and the China Postdoctoral Science Foundation (CPSF).
\end{acknowledgments}

\end{document}